\documentstyle[12pt,epsf]{article}

\newcommand{\beq}{\begin{equation}}
\newcommand{\eeq}{\end{equation}}
\newcommand{\beqa}{\begin{eqnarray}}
\newcommand{\eeqa}{\end{eqnarray}}
\newcommand{\ba}{\begin{array}}
\newcommand{\ea}{\end{array}}

\newcommand{\al}{\alpha}
\newcommand{\as}{\alpha_s}
\newcommand{\api}{\frac{\alpha_s}{\pi}}
\newcommand{\msbar}{\overline{\mbox{MS}}}

\newcommand{\scsi}{\scriptsize}

\newcommand{\ep}{\epsilon}
\newcommand{\drho}{\delta\rho}
\newcommand{\mbar}{\bar{m}_t}

\newcommand{\sineff}{\sin^2\!\bar{\Theta}}
\newcommand{\dr}{\Delta r}
\newcommand{\drtb}{\Delta r_{tb}}
\newcommand{\drtbbar}{\Delta \bar{r}_{tb}}
\newcommand{\dktb}{\Delta\kappa_{tb}}
\newcommand{\ga}{\gamma}
\newcommand{\mzmt}{\frac{M_Z^2}{M_t^2}}

\newcommand{\lmzmt}{\,l_Z}
\newcommand{\lmzmtz}{\,l_Z^{\,\,2}}
\newcommand{\lms}{\bar{l}_Z}
\newcommand{\lmsz}{\bar{l}_Z^{\,\,2}}

\begin{document}

\begin{titlepage}
\noindent
%
%
\hfill SLAC-PUB-95-6851  \\
\mbox{}
\hfill TTP95--13\footnote{The complete postscript file of this
preprint, including figures, is available via anonymous ftp at
ttpux2.physik.uni-karlsruhe.de (129.13.102.139) as
/ttp95-13/ttp95-13.ps or via www at
http://ttpux2.physik.uni-karlsruhe.de/cgi-bin/preprints/
Report-no: TTP95-13}\\
\mbox{}
\hfill hep-ph/9504413\\

\vspace{.5cm}

\begin{center}
\begin{LARGE}
 {\bf QCD Corrections
      from Top Quark to  \\[.00em]
      Relations between Electroweak Parameters \\[.2em]
      to Order $\as^2$}\footnote{Work supported
                                 in part by the Department of Energy,
                                 Contract DE-ACO3-76SF00515
                                 and by DFG under Contract Ku 502/6-1.}
\end{LARGE}

\vspace{.8cm}

\begin{large}
 K.G.~Chetyrkin$^{a,b}$,
 J.H.~K\"uhn$^{b,c}$,
 M.~Steinhauser$^{b}$
\end{large}

\vspace{.5cm}
\begin{itemize}
\item[$^a$]
   Institute for Nuclear Research\\
   Russian Academy of Sciences, 60th October Anniversary Prospect 7a,
   Moscow, 117312, Russia
\item[$^b$]
   Institut f\"ur Theoretische Teilchenphysik\\
   Universit\"at Karlsruhe, Kaiserstr. 12,    Postfach 6980,
   D-76128 Karlsruhe, Germany\\
\item[$^c$]
   Stanford Linear Accelerator Center\\
   Stanford University, Stanford, CA 94309
\end{itemize}

\vspace{.5cm}

\begin{abstract}
\noindent
The vacuum polarization functions $\Pi(q^2)$ of charged and neutral
gauge bosons which arise from top and bottom quark loops lead to
important shifts in relations between electroweak parameters
which can be measured with ever-increasing precision. The large
mass of the top quark allows approximation of these functions through
the first two terms of an expansion in $M_Z^2/M_t^2$.
The first three terms of the Taylor series of $\Pi(q^2)$ are
evaluated analytically up to order $\as^2$.
The first two are required to derive the approximation, the
third can be used to demonstrate the smallness of the neglected terms.
The paper
improves earlier results
based on the leading term $\propto G_F M_t^2 \as^2$.
Results for the subleading
contributions to $\dr$ and the effective mixing angle $\sineff$ are presented.
\end{abstract}

\centerline{(Submitted to Physical Review Letters)}


\vfill
\end{center}
\end{titlepage}

\renewcommand{\arraystretch}{2}

\section{Introduction}

The precision achieved in recent measurements of electroweak observables
\cite{Sch94}
has surpassed by far earlier expectations.
The predictions for these quantities and the relations between them, which
are based on the Standard Model (SM),
are strongly affected by radiative corrections.
Particularly important are those induced by virtual contributions from the
heavy top quark
\cite{Vel77}.
Their verification provides an important test of the theory and its quantum
corrections. The agreement between
the value of $M_t$ suggested by the CDF
and D0 collaborations
\cite{CDF}
of $M_t=176\pm8\pm10$ GeV and $M_t=199^{+19}_{-21}\pm22$ GeV, respectively,
and deduced indirectly from precision measurements
of $M_t=173^{+12+18}_{-13-20}$ GeV
\cite{Sch94},
constitutes a triumph of the SM and a verification of the
quantum corrections with increasing precision.
A more refined understanding of these effects
is on the agenda, including two- and even the dominant three-loop contributions
\cite{AvdFleMikTar94,CheKueSte95}.

As far as contributions from the top-bottom multiplet are concerned, the
perturbative results for the self-energies
$\Pi_{WW}(q^2), \Pi_{ZZ}(q^2), \Pi_{\ga\ga}(q^2)$ and $\Pi_{\ga Z}(q^2)$
are available in the literature for arbitrary top and bottom masses
up to order $\as$
\cite{DjoVer87Djo88,Kni90,KniKueStu88}.

This allows evaluation not only of the leading corrections, which are
governed by the $\rho$ parameter and increase with $M_t^2$,
but also of
the subleading terms.
These are required for a complete calculation of $\dr$ (entering the
relation between $G_F, M_W^2, M_Z^2$ and $\al$) or of the effective mixing
angle $\sineff$ (governing asymmetries in $Z$ production and decay).
Recently also three loop QCD corrections to the $\rho$ parameter
of ${\cal O}(G_F M_t^2 \as^2)$ have been calculated
\cite{AvdFleMikTar94,CheKueSte95}.
These, in turn, control the dominant terms of order
$G_F M_t^2 \as^2$ in $\dr$ and $\sineff$.

The technique described in
\cite{AvdFleMikTar94,CheKueSte95}
can be employed to obtain also the Taylor series
coefficients of $\Pi(q^2)$ around $q^2=0$, in principle to arbitrary
orders in $q^2$ and in second order in $\as$.
It will be demonstrated below that the two lowest terms in the expansion
of $\dr$ in $M_Z^2/M_t^2$ provide an excellent approximation to the full
answer in one- and two-loop approximation, corresponding to
$\as^0$ and $\as^1$ corrections. This justifies the expectation that
also in order $\as^2$ the leading terms $\propto G_F M_t^2 \as^2$ plus the
subleading terms provide an adequate description of the complete
answer for $\dr$ and $\sineff$.
This is verified by calculating the terms $\propto M_Z^2/M_t^2$ which
indeed turn out to be negligible. The complete result for $\Delta r$ and
$\sineff$ to order $\as^2$ is therefore at hand.

\section{The $M_W$ -- $M_Z$ connection and the effective mixing angle}

It has become customary to express the magnitude of radiative
corrections in the relation between $M_W, M_Z, G_F$ and $\al$ through
the quantity $\dr$, defined through
\cite{Sir80}
\beq
M_W^2 = \frac{M_Z^2}{2} \left(
        1+\sqrt{1-\frac{4\pi\al}{\sqrt{2}M_Z^2 G_F(1-\dr)}}
                          \,\,\right).
\label{mweqn}
\eeq
The influence of a heavy quark doublet can be expressed through
the transversal parts of the gauge boson self energies:
\beq
\drtb = \frac{c^2}{s^2}\mbox{Re}
     \left(\frac{\Pi_{WW}(M_W^2)}{M_W^2}-\frac{\Pi_{ZZ}(M_Z^2)}{M_Z^2}\right)
      + \tilde{\Pi}_{\ga\ga}(0)
      + \frac{1}{M_W^2}\left(\Pi_{WW}(0)-\mbox{Re}\Pi_{WW}(M_W^2)\right)
\eeq
where $\tilde{\Pi}_{\ga\ga}(q^2)=\Pi_{\ga\ga}(q^2)/q^2$.
In terms of the transversal parts of vector and axial current correlators
$\Pi^V$ and $\Pi^A$ the building blocks for $\dr$ are given by
\beqa
\Pi_{WW} &=& \frac{g^2}{8} \Big[\Pi^V(q^2,m_t,m_b) + \Pi^A(q^2,m_t,m_b) \Big],
\nonumber\\
\Pi_{ZZ} &=& \frac{g^2}{16c^2}
             \sum_{i=t,b}\Big[v_i^2\Pi^{V,i}(q^2,m_i)+\Pi^{A,i}(q^2,m_i)\Big]
             + \frac{g^2}{16c^2} \Pi^{A,S}(q^2,m_t,m_b),
\nonumber\\
\Pi_{\ga\ga} &=& g^2 s^2 \sum_{i=t,b} Q_i^2\Pi^{V,i}(q^2,m_i),
\label{polfun}\\
\Pi_{\ga Z} &=& -\frac{g^2 s}{4c} \sum_{i=t,b} Q_i v_i \Pi^{V,i}(q^2,m_i),
\nonumber
\eeqa
with $v_i=2 I_3^i - 4 s^2 Q_i$. The sin and cos of the weak mixing angle
are denoted by s and c. The ``singlet'' contribution to the axial
current correlator $\Pi^{A,S}$ which originates from double triangle
diagrams
occurs first in order $\as^2$ and has been displayed separately.

Throughout this paper the mass of the bottom quark will be neglected. To
circumvent the mass singularity which originates from the bottom
loop contribution $\tilde{\Pi}_{\ga\ga}(0)$ is replaced by
$\tilde{\Pi}_{\ga\ga}(q^2=M_Z^2)$.
The difference is accounted for by dispersion relations with input
from the actual measurement of $\sigma(e^+ e^- \to hadrons)$ in
the low energy region
\cite{Jeg}.

The evaluation of $\Pi(q^2)$ up to three-loop diagrams involving massless
diagrams only is performed with the FORM
\cite{Ver91}
package MINCER
\cite{LarTkaVer91}
implementing an algorithm developed in
\cite{CheTka81}.
For diagrams involving a heavy top quark the approximation based on a
Taylor series of $\Pi(q^2)$ around $q^2=0$, which is equivalent to an
expansion in $M_Z^2/M_t^2$, leads to an adequate approximation:
\beqa
\drtbbar & \equiv & \drtb - \tilde{\Pi}_{\ga\ga}(0)
                          + \mbox{Re}\tilde{\Pi}_{\ga\ga}(M_Z^2)
\nonumber\\
         &=& \frac{c^2}{s^2}
   \left(\frac{\Pi_{WW}(0)}{M_W^2}-\frac{\Pi_{ZZ}(0)}{M_Z^2}\right)
\label{deltar}\\
&&
  +\frac{c^2}{s^2}\left(\Pi_{WW}^{\prime}(0)-\Pi_{ZZ}^{\prime}(0)|_t
                      -\mbox{Re}\frac{\Pi_{ZZ}(M_Z^2)|_b}{M_Z^2}\right)
\nonumber\\
&&
\hphantom{\hspace{12em}}
  +\tilde{\Pi}_{\ga\ga}(0)|_t + \mbox{Re}\tilde{\Pi}_{\ga\ga}(M_Z^2)|_b
  -\Pi_{WW}^{\prime}(0)
\nonumber\\
&&
  +\frac{c^2}{s^2}\left(\Pi_{WW}^{\prime\prime}(0)\frac{M_W^2}{2}
                       -\Pi_{ZZ}^{\prime\prime}(0)|_t\frac{M_Z^2}{2}\right)
  +\tilde{\Pi}^{\prime}_{\ga\ga}(0)|_t M_Z^2
  -\Pi_{WW}^{\prime\prime}(0)\frac{M_W^2}{2}
\nonumber\\
&& +\ldots
\nonumber
\eeqa
The first line in this expansion is determined by the $\rho$ parameter,
leads to terms $\propto G_F M_t^2$, and has been recently calculated up
to order $\as^2$.

The second line leads to terms of order $G_F M_Z^2$ and
$G_F M_Z^2 \ln M_t^2/M_Z^2$ and will be calculated below up to order
$\as^2$. The subsequent terms are of order $G_F M_Z^4/M_t^2$.
Even after charge and mass renormalization $\Pi(q^2)$
exhibits (in dimensional regularization) $1/\ep$
singularities which cancel in the combination (\ref{deltar}).

The reliability of the $M_Z^2/M_t^2$ expansion is
demonstrated in Fig. \ref{figdr} for the one- and two-loop
results respectively.
The full answer (solid line) is compared to the approximation based on
the term quadratic im $M_t$ (dotted line) and the approximation including
constant plus $\ln M_Z^2/M_t^2$ terms (dashed line).
The latter provides an excellent approximation in the range
$M_t>150$ GeV; corrections of order $G_F M_Z^4/M_t^2$ may safely be
ignored (dash-dotted line).

The leading terms of the expansion are given by
\cite{HalKniSto93} ($X_t = G_F M_t^2/(8\sqrt{2}\pi^2)$)
\beqa
\drtbbar & = & -3 X_t\frac{c^2}{s^2} \Bigg\{
                1 + \api C_F \left(  - \frac{1}{2} - \zeta(2) \right)
      +\frac{M_Z^2}{M_t^2}\Bigg[
           \ln\frac{M_Z^2}{M_t^2}\left(  -\frac{2}{3} + \frac{8}{9}s^2 \right)
          +\frac{1}{3} - \frac{16}{27} s^2
\nonumber\\
&&
          +\api C_F\Bigg(\ln\frac{M_Z^2}{M_t^2}
                            \left(-\frac{1}{2} + \frac{2}{3}s^2 \right)
                 - 2 \zeta(3) + \frac{4}{9}\zeta(2) +\frac{1}{2}
\nonumber\\
&&
                 + s^2\left(\frac{8}{3}\zeta(3)
                           - \frac{8}{9}\zeta(2) - \frac{5}{3}\right)
                   \Bigg)
                        \Bigg]
      +\frac{M_Z^4}{M_t^4}\Bigg[
                 - \frac{2}{5} + \frac{103}{90} s^2 - s^4
\nonumber\\
&&
                 +\api C_F  \left(-\frac{523}{3240} + \frac{3}{2} s^2 \zeta(2)
                                  + \frac{827}{3888} s^2 - s^4 \zeta(2)
                                  - \frac{25}{24} s^4 - \frac{1}{2} \zeta(2)
                            \right)
                        \Bigg]
                                     \Bigg\}.
\label{delr}
\eeqa
The full result is given e.g. in
\cite{Kni90,HalKniSto93}.

A second quantity of practical interest is the effective weak mixing
angle, which governs in particular the asymmetries
\cite{Hol}
in $Z$ boson production and decay. It is related to
$\sin^2 \Theta \equiv 1-M_W^2/M_Z^2$ through a correction factor
\beq
\sineff = (1+\Delta\kappa)\sin^2\Theta
\eeq
which in turn is influenced by the polarization functions of Eq. (\ref{polfun})
\beqa
\dktb &=& -\frac{c}{s}\frac{\Pi_{\ga Z}(M_Z^2)}{M_Z^2}
     -\frac{c^2}{s^2}\mbox{Re}
     \left(\frac{\Pi_{WW}(M_W^2)}{M_W^2}-\frac{\Pi_{ZZ}(M_Z^2)}{M_Z^2}\right)
\nonumber
\\
&=&
         -3 X_t\frac{c^2}{s^2} \Bigg\{
                -1 + \api C_F \left(   \frac{1}{2} + \zeta(2) \right)
      +\frac{M_Z^2}{M_t^2}\Bigg[
           \ln\frac{M_Z^2}{M_t^2}\left(  \frac{2}{3} - \frac{4}{9}s^2 \right)
          -\frac{1}{3} + \frac{8}{27} s^2
\nonumber\\
&&
          +\api C_F\Bigg(\ln\frac{M_Z^2}{M_t^2}
                            \left(\frac{1}{2} - \frac{1}{3}s^2 \right)
                 + 2 \zeta(3) - \frac{4}{9}\zeta(2) -\frac{1}{2}
\nonumber\\
&&
                 + s^2\left(-\frac{4}{3}\zeta(3)
                           + \frac{4}{9}\zeta(2) + \frac{5}{6}\right)
                   \Bigg)
                        \Bigg]
      +\frac{M_Z^4}{M_t^4}\Bigg[
                  \frac{2}{5} - \frac{37}{45} s^2 + \frac{1}{2}s^4
\nonumber\\
&&
                 +\api C_F  \left(\frac{523}{3240} - s^2 \zeta(2)
                                  - \frac{713}{1944} s^2
                                  +\frac{1}{2} s^4 \zeta(2)
                                  + \frac{25}{48} s^4 + \frac{1}{2} \zeta(2)
                            \right)
                        \Bigg]
\nonumber\\
&&
     + i \pi\frac{M_Z^2}{M_t^2}\Bigg[
            -\frac{4}{9} s^2 + \frac{16}{27} s^4
   + \api C_F \Bigg(-\frac{1}{3} s^2 + \frac{4}{9} s^4\Bigg)
            \Bigg]
                                     \Bigg\}.
\label{delkap}
\eeqa
The full analytic result and the approximation based on the $M_Z/M_t^2$
expansion are compared in
Fig. \ref{figdk}, adopting the same notation
as in Fig. \ref{figdr}.
{}From these figures and from Eqs. (\ref{delr},\ref{delkap}) it is evident that
(for $M_t\approx 180$ GeV) the next-to-leading corrections amount
to about 25\% of the $G_F M_t^2$ terms.
The next-to-next-to-leading terms of order $G_F M_Z^2/M_t^2$ are below
1.2\%. These considerations justify the restriction of $\as^2$ corrections
to the first two or at most three terms in the $M_Z^2/M_t^2$ expansion.

\section{Three-Loop Corrections}

In addition to the massless diagrams there are two further types of
integrals:
The evaluation of the
derivatives of $\Pi(q^2)$ resulting from diagrams where the massive top
quark is coupled to the $W$ or $Z$ is reduced to the evaluation of
tadpole integrals discussed in
\cite{AvdFleMikTar94,CheKueSte95}.
The derivatives are obtained through Taylor expansion of the respective
integrands up to ${\cal O}(q^2)$
\beq
\Pi_{\mu\nu}(q) = g_{\mu\nu}\Pi(q^2) + q_\mu q_\nu\Pi_2(q^2)
\eeq
and projecting out the transverse part.
Following
\cite{CheTka81,Bro92}
the resulting tadpole integrals are
subsequently reduced to the master set listed in
\cite{AvdFleMikTar94,CheKueSte95}.
Diagrams involving external massless quark loops and internal top loops
(and the anomaly graph) are treated with the large mass expansion
technique
\cite{Smi94}.

Setting $C_A=3, C_F=4/3$ and $\mu^2=\mbar^2$
a fairly compact form for the subleading parts of $\drtbbar$
and $\dktb$ is obtained ($x_t = G_F \mbar^2/(8\sqrt{2}\pi^2)
        ,\lms=\ln M_Z^2/\mbar^2$):

\beqa
\lefteqn{\drtbbar^{\scsi\msbar}  =  -\frac{c^2}{s^2}
          \drho_{\scsi\msbar} - 3 \frac{c^2}{s^2} x_t \frac{M_Z^2}{\mbar^2}}
\\
&&
\Bigg\{
        \frac{1}{3} + \frac{8}{9} s^2 \lms - \frac{16}{27} s^2
       -\frac{2}{3} \lms
\nonumber\\
&&
       + \frac{\as}{4\pi}\Bigg(
           \frac{88}{9} + \frac{128}{9} s^2 \zeta(3)
         - \frac{128}{27} s^2 \zeta(2)
         + \frac{32}{9} s^2 \lms - \frac{496}{27} s^2 - \frac{32}{3} \zeta(3)
         + \frac{64}{27} \zeta(2) - \frac{8}{3} \lms \Bigg)
\nonumber\\
&&
       +\left(\frac{\as}{4\pi}\right)^2 \Bigg[
         n_f \Bigg(  - \frac{4480}{243} + \frac{256}{27} s^2 \zeta(3) \lms
                     - \frac{2944}{81} s^2 \zeta(3)
                     + \frac{448}{27} s^2 \zeta(2)
                     - \frac{352}{27} s^2 \lms
                     + \frac{32}{27} s^2 \lmsz
\nonumber\\
&&
\hphantom{\frac{\as}{4\pi} }
          + \frac{2696}{243} s^2
          - \frac{64}{9} \zeta(3) \lms + \frac{2080}{81} \zeta(3)
          - \frac{176}{27} \zeta(2) + \frac{88}{9} \lms
          - \frac{8}{9} \lmsz
          \Bigg)
\nonumber\\
&&
\hphantom{\frac{\as}{4\pi} }
         + \frac{94957}{243} + 856 S_2 s^2 - 428 S_2 + \frac{256}{81} D_3 s^2
         - \frac{128}{81}D_3 - \frac{4480}{27} s^2 \zeta(3) \lms
         + \frac{12016}{27} s^2 \zeta(3)
\nonumber\\
&&
\hphantom{\frac{\as}{4\pi} }
         + \frac{12032}{81} s^2 \zeta(4)
         - \frac{3200}{27} s^2 \zeta(5) - \frac{52816}{243} s^2 \zeta(2)
         - \frac{256}{27} s^2 B_4
         + \frac{688}{3} s^2 \lms
         - \frac{560}{27} s^2 \lmsz
\nonumber\\
&&
\hphantom{\frac{\as}{4\pi} }
          - \frac{24164}{81} s^2
         + \frac{1120}{9} \zeta(3) \lms
         - \frac{8117}{18} \zeta(3) - \frac{6016}{81} \zeta(4)
         + \frac{800}{9} \zeta(5)
\nonumber\\
&&
\hphantom{\frac{\as}{4\pi} }
         + \frac{22736}{243} \zeta(2)
         + \frac{128}{27} B_4
         - \frac{1400}{9} \lms + \frac{116}{9} \lmsz
\Bigg]
\Bigg\},
\nonumber\\
\lefteqn{\dktb^{\scsi\msbar} =  \frac{c^2}{s^2}
          \drho_{\scsi\msbar} - 3 \frac{c^2}{s^2} x_t \frac{M_Z^2}{\mbar^2}}
\\
&&
\Bigg\{
       - \frac{1}{3} - \frac{4}{9} s^2 \lms + \frac{8}{27} s^2
       + \frac{2}{3} \lms
\nonumber\\
&&
       + \frac{\as}{4\pi}\Bigg(
                   - \frac{88}{9} - \frac{64}{9} s^2 \zeta(3)
                   + \frac{64}{27} s^2 \zeta(2)
                   - \frac{16}{9} s^2 \lms + \frac{248}{27} s^2
                   + \frac{32}{3} \zeta(3) - \frac{64}{27} \zeta(2)
                   + \frac{8}{3} \lms \Bigg)
\nonumber\\
&&
       +\left(\frac{\as}{4\pi}\right)^2 \Bigg[
         n_f  \Bigg( \frac{4480}{243} - \frac{128}{27} s^2 \zeta(3) \lms
                   + \frac{1472}{81} s^2 \zeta(3)
                   - \frac{224}{27} s^2 \zeta(2) + \frac{176}{27} s^2 \lms
\nonumber\\
&&
\hphantom{\frac{\as}{4\pi} }
                   - \frac{16}{27} s^2 \lmsz - \frac{1348}{243} s^2
                   + \frac{64}{9} \zeta(3) \lms - \frac{2080}{81} \zeta(3)
                   + \frac{176}{27} \zeta(2) - \frac{88}{9} \lms
                   + \frac{8}{9} \lmsz \Bigg)
\nonumber\\
&&
\hphantom{\frac{\as}{4\pi} }
                 - \frac{94957}{243} - 428 S_2 s^2 + 428 S_2
                 - \frac{128}{81} D_3 s^2 + \frac{128}{81} D_3
                 + \frac{2240}{27} s^2 \zeta(3) \lms
\nonumber\\
&&
\hphantom{\frac{\as}{4\pi} }
                 - \frac{6008}{27} s^2 \zeta(3) - \frac{6016}{81} s^2 \zeta(4)
                 + \frac{1600}{27} s^2 \zeta(5)
                 + \frac{26408}{243} s^2 \zeta(2) + \frac{128}{27} s^2 B_4
                 - \frac{344}{3} s^2 \lms
\nonumber\\
&&
\hphantom{\frac{\as}{4\pi} }
                 + \frac{280}{27} s^2 \lmsz
                 + \frac{12082}{81} s^2
                 - \frac{1120}{9} \zeta(3) \lms
                 + \frac{8117}{18} \zeta(3)
                 + \frac{6016}{81} \zeta(4)
\nonumber\\
&&
\hphantom{\frac{\as}{4\pi} }
                 - \frac{800}{9} \zeta(5)
                 - \frac{22736}{243} \zeta(2)
                 - \frac{128}{27} B_4
                 + \frac{1400}{9} \lms - \frac{116}{9} \lmsz
\Bigg]
\nonumber\\
&&
     + i \pi\Bigg[
            -\frac{4}{9} s^2 + \frac{16}{27} s^4
   + \frac{\as}{4\pi} \Bigg(-\frac{16}{9} s^2 + \frac{64}{27} s^4\Bigg)
\nonumber\\
&&
\hphantom{\frac{\as}{4\pi} }
   + \left(\frac{\as}{4\pi}\right)^2  \Bigg(
            n_f \bigg( \frac{176}{27}s^2 - \frac{32}{27}s^2 \lms
                    - \frac{128}{27}s^2\zeta(3)
               -\frac{704}{81}s^4 + \frac{128}{81}s^4 \lms
               + \frac{512}{81}s^4 \zeta(3) \bigg)
\nonumber\\
&&
\hphantom{\frac{\as}{4\pi}\frac{\as}{4\pi} }
    -\frac{344}{3}s^2 + \frac{560}{27}s^2 \lms + \frac{2240}{27}s^2\zeta(3)
    +\frac{1376}{9}s^4 - \frac{2240}{81}s^4 \lms - \frac{8960}{81}s^4\zeta(3)
             \Bigg)
             \Bigg]
\Bigg\}
\nonumber
\eeqa
with
\cite{AvdFleMikTar94}
\begin{eqnarray*}
B_4 &=& 16\, \mbox{Li}_4\left(\frac{1}{2}\right)
      + \frac{2}{3}\log^4 2
      - \frac{2}{3}\pi^2\log^2 2
      - \frac{13}{180}\pi^4
     = -1.76280\ldots \\
S_2 &=& \frac{4}{9\sqrt{3}} \mbox{Cl}_2\left(\frac{\pi}{3}\right)
     = 0.260434\ldots\\
D_3 &=& -3.02700\ldots.
\end{eqnarray*}
The formula for $\delta\rho_{\scsi\msbar}$ can be found in
\cite{CheKueSte95}.

This result is formulated in terms of the $\msbar$ coupling
$\as$ and the mass $\mbar$.
Employing the two-loop relation between the
$\msbar$ mass and the pole mass $M_t$
\cite{BroGraGraSch90}
the $\msbar$ results are easily expressed in terms of $M_t$.
The final result, after setting $n_f=6$, reads ($\mu^2=M_t^2$):
\beqa
\lefteqn{\drtbbar^{OS} = -\frac{c^2}{s^2} 3 X_t }
&&
\hphantom{\dktb^{OS}=\frac{c^2}{s^2}X_t}
\Bigg\{
        1 + \mzmt \Big(0.3333 - 0.6667 \lmzmt +
        (-0.5926 + 0.8889 \lmzmt) s^2 \Big)
\nonumber\\
&&
\hphantom{\frac{\as}{4\pi} }
        +\left(\mzmt\right)^2
         \Big(-0.4 + 1.144 s^2  - s^4\Big)
\nonumber\\
&&
    +\api \Big[-2.8599 + \mzmt \Big(-1.5640 - 0.6667 \lmzmt +
          (0.1022 + 0.8889 \lmzmt) s^2 \Big)
\nonumber\\
&&
\hphantom{\frac{\as}{4\pi} }
         +\left(\mzmt\right)^2
                     \Big(-1.312 + 3.573 s^2  - 3.582 s^4 \Big)
          \Big]
\nonumber\\
&&
    +\left(\api\right)^2
         \Big[-14.594 + \mzmt \Big(-17.224 + 0.08829 \lmzmt
               + 0.4722 \lmzmtz
\nonumber\\
&&
\hphantom{\frac{\as}{4\pi} }
         + (22.6367 + 1.2527 \lmzmt - 0.8519 \lmzmtz ) s^2 \Big)
         + \left(\mzmt\right)^2
            \Big(-7.7781 - 0.07226 \lmzmt
\nonumber\\
&&
\hphantom{\frac{\as}{4\pi} }
                 + 0.004938 \lmzmtz
         +(21.497 + 0.05794 \lmzmt - 0.006584 \lmzmtz ) s^2  - 21.0799 s^4
                   \Big)
\Big]
\Bigg\}
\nonumber\\
\lefteqn{\dktb^{OS}  = -\frac{c^2}{s^2} 3 X_t }
&&
\hphantom{\dktb^{OS}=\frac{c^2}{s^2} X_t }
\Bigg\{
     -1 + \mzmt \Big(-0.3333 + 0.6667 \lmzmt +
       (0.2963 - 0.4444 \lmzmt) s^2 \Big)
\nonumber\\
&&
\hphantom{\frac{\as}{4\pi} }
        +\left(\mzmt\right)^2
         \Big(0.4 - 0.8222 s^2  + 0.5 s^4\Big)
\nonumber\\
&&
    +\api \Big[2.8599 + \mzmt \Big(1.564 + 0.6667 \lmzmt +
          (-0.051103 - 0.4444 \lmzmt) s^2 \Big)
\nonumber\\
&&
\hphantom{\frac{\as}{4\pi} }
       \left(\mzmt\right)^2  \Big(1.312 - 2.682 s^2  + 1.791 s^4 \Big)
\Big]
\nonumber\\
&&
    +\left(\api\right)^2
      \Big[14.594 + \mzmt \Big(17.224 - 0.08829 \lmzmt - 0.47222 \lmzmtz
\nonumber\\
&&
\hphantom{\frac{\as}{4\pi} }
       +   (-11.3184 - 0.6263 \lmzmt + 0.4259 \lmzmtz ) s^2 \Big)
       + \left(\mzmt\right)^2 \Big(7.7781 + 0.072263 \lmzmt
\nonumber\\
&&
\hphantom{\frac{\as}{4\pi} }
                              - 0.004938 \lmzmtz
       +(-16.0186 - 0.02897 \lmzmt + 0.003292 \lmzmtz ) s^2  + 10.54 s^4 \Big)
\Big]
\nonumber\\
&&
        +i \mzmt \Bigg[
              -1.396 s^2  + 1.862 s^4
        + \api  (-1.396 s^2  + 1.862 s^4 )
\nonumber\\
&&
\hphantom{\frac{\as}{4\pi} }
       + \left(\api\right)^2  \Big[(-1.968 + 2.676 \lmzmt) s^2
          +(2.6235 - 3.5682 \lmzmt) s^4
\nonumber\\
&&
\hphantom{\frac{\as}{4\pi}\frac{\as}{4\pi} }
       +  \mzmt  \Big((-0.09102 + 0.02069 \lmzmt) s^2
                             + (0.1214 - 0.02758 \lmzmt) s^4 \Big)\Big]
                 \Bigg]
\Bigg\},
\nonumber
\eeqa
where $\lmzmt=\ln M_Z^2/M_t^2$.
In these formulae also the terms of order $G_F\as^2 M_Z^4/M_t^2$ are given.
As shown in Fig. \ref{figdr} and \ref{figdk} and Table \ref{tab1} their
effect of the numerical result is extremely small and can safely
be neglected.

\begin{table}[th]
\renewcommand{\arraystretch}{1.3}
\begin{center}

\begin{tabular}{|l||r|r|r|}
\hline
     & $M_t^2$
     & $M_t^2$ + const.
     & $M_t^2$ + const. +$1/M_t^2$           \\
\hline
\hline
$\delta M_W/M_W$ (OS)       & 0.00677& 0.00838 & 0.00825\\
$\delta\sineff/\sineff$ (OS) & -0.01618& -0.01832 & -0.01814\\
\hline
%
%
\hline
$\delta M_W/M_W$ ($\msbar$)       & 0.00674& 0.00833 & 0.00821\\
$\delta\sineff/\sineff$ ($\msbar$)& -0.01610& -0.01823 & -0.01804\\
\hline
\end{tabular}

\end{center}
\caption{\label{tab1}Numerical results for the $M_t^2$, the constant plus
                     logarithmic and the $1/M_t^2$ contributions.
         For the numerical evaluation of $\delta\sineff/\sineff$ only the
         real part of $\dktb$ is taken.}
\end{table}

\begin{table}[th]
\renewcommand{\arraystretch}{1.3}
\begin{center}

\begin{tabular}{|l||r|r|r|}
\hline
$\delta M_W$ in MeV  & $\alpha_s^0$ & $\alpha_s^1$ & $\alpha_s^2$ \\
\hline
\hline
$M_t^2$             & 611.9 & -61.3 & -10.9 \\
const.              & 136.6 & -6.0  & -2.6  \\
$1/M_t^2$           & -9.0  & -1.0  & -0.2  \\
\hline
\end{tabular}

\end{center}
\caption{\label{tab2} The change in $M_W$ separated according
                      to powers in $\alpha_s$ and $M_t$ in the on-shell
                      scheme.}
\end{table}

In the lowest diagrams of Fig. \ref{figdr} and \ref{figdk}
the ${\cal O}(\as^2)$ corrections of $\drtbbar$ and $\dktb$ are presented
as functions of $M_t$ and
the quality of the $M_Z^2/M_t^2$ expansion is confirmed.
The difference between the quadratic term (dotted line) and the
constant plus log term (dashed line) amounts to about 25\%. Adding the
subsequent term proportional $M_Z^2/M_t^2$ (dashed dotted line) barely
affects the answer.
The numerical effects on $M_W$ and $\sineff$ are given in
Table \ref{tab1}. The numbers are obtained with the following input data:
$\as(M_t^2)=0.1092$ (corresponding $\as^{(5)}(M_Z^2)=0.120$),
$M_t=175$ GeV, $M_Z=91.188$ GeV, $M_H=300$ GeV,
$\al=1/137.04$ and
$G_F=1.16639\, 10^{-5}$ GeV$^{-2}$.
In each column the terms of order $\as^0$, $\as^1$ and $\as^2$
belonging to the corresponding expansion in the top mass
are added up.
For $\Delta r$ in Eq. (\ref{mweqn}) we used
$\Delta r = \Delta\alpha+\drtbbar+\delta_{rem}$
with $\Delta\alpha=0.05940$ \cite{Jeg}.
$\delta_{rem}$ contains all contributions
of order $\alpha$ which are not present in the other two
pieces and can e.g. be found in
\cite{BurJeg90}.
In Table \ref{tab2} the contributions are listed separately  according to
powers of $\alpha_s$ and $M_t$.
One observes that
the absolute prediction for the $W$ mass
is changed by $-10.9$ MeV if the  $\as^2 M_t^2$ term is added to
the full two-loop result.
This increases to $-13.7$ MeV if
also the constant and the $1/M_t^2$ suppressed terms terms are added.
(For a fictitious top mass of 100 GeV the numbers would be $-4.2$ MeV
and $-7.9$ MeV respectively.)

Summary: Top mass dependent corrections to relations between electroweak
parameters have been calculated up to order $\as^2$, with the help
of an expansion in $M_Z^2/M_t^2$. The quality of the approximation
has been confirmed in one-, two- and three-loop approximations.

\vspace{5ex}
{\bf Acknowledgments}

\noindent
We would like to thank W. Hollik and M. Peskin for valuable discussions.
J.K. would like to thank the SLAC theory group for hospitality and the
Volkswagen-Stiftung grant I/70 452 for generous support.

%
%

\begin{figure}[b]
\centerline{\bf Figure Captions}

\caption{\label{figdr}$\drtbbar$ as a function of $M_t$. The dotted, dashed
         and dash-dotted curves correspond to an increasing number of
         terms in the approximation. In the figures a value $s^2=0.2321$
         is chosen.}
\caption{\label{figdk}$\Delta\kappa_{tb}$ as a function of $M_t$.
         The conventions are the same as for $\drtbbar$. Here only the
         real part is plotted.}

\end{figure}

\setcounter{figure}{0}

\begin{figure}[ht]
 \begin{center}
 \begin{tabular}{c}
   \epsfxsize=11.5cm
   \leavevmode
   \epsffile[50 290 520 560]{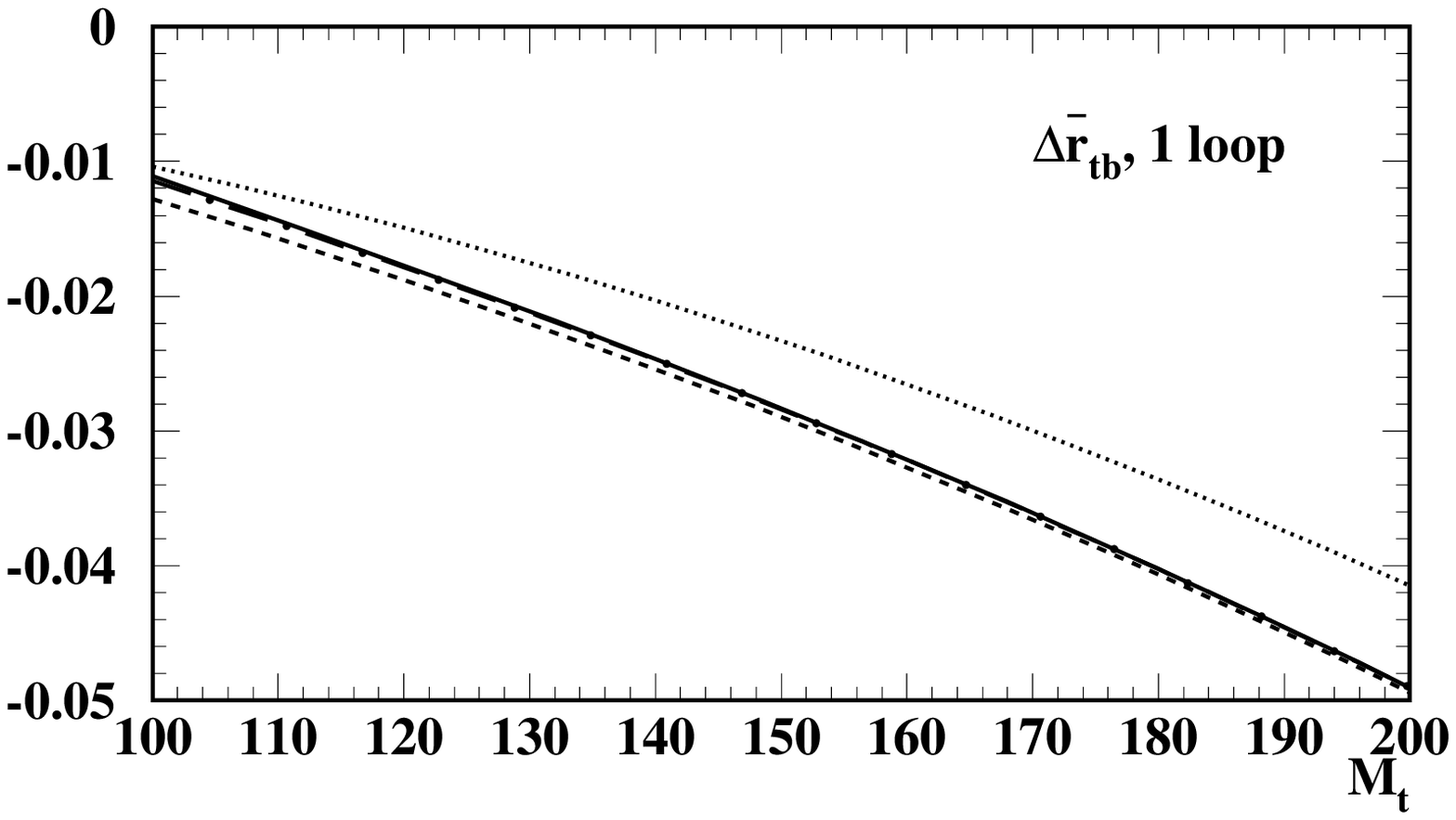} \\
   \epsfxsize=11.5cm
   \leavevmode
   \epsffile[50 290 520 560]{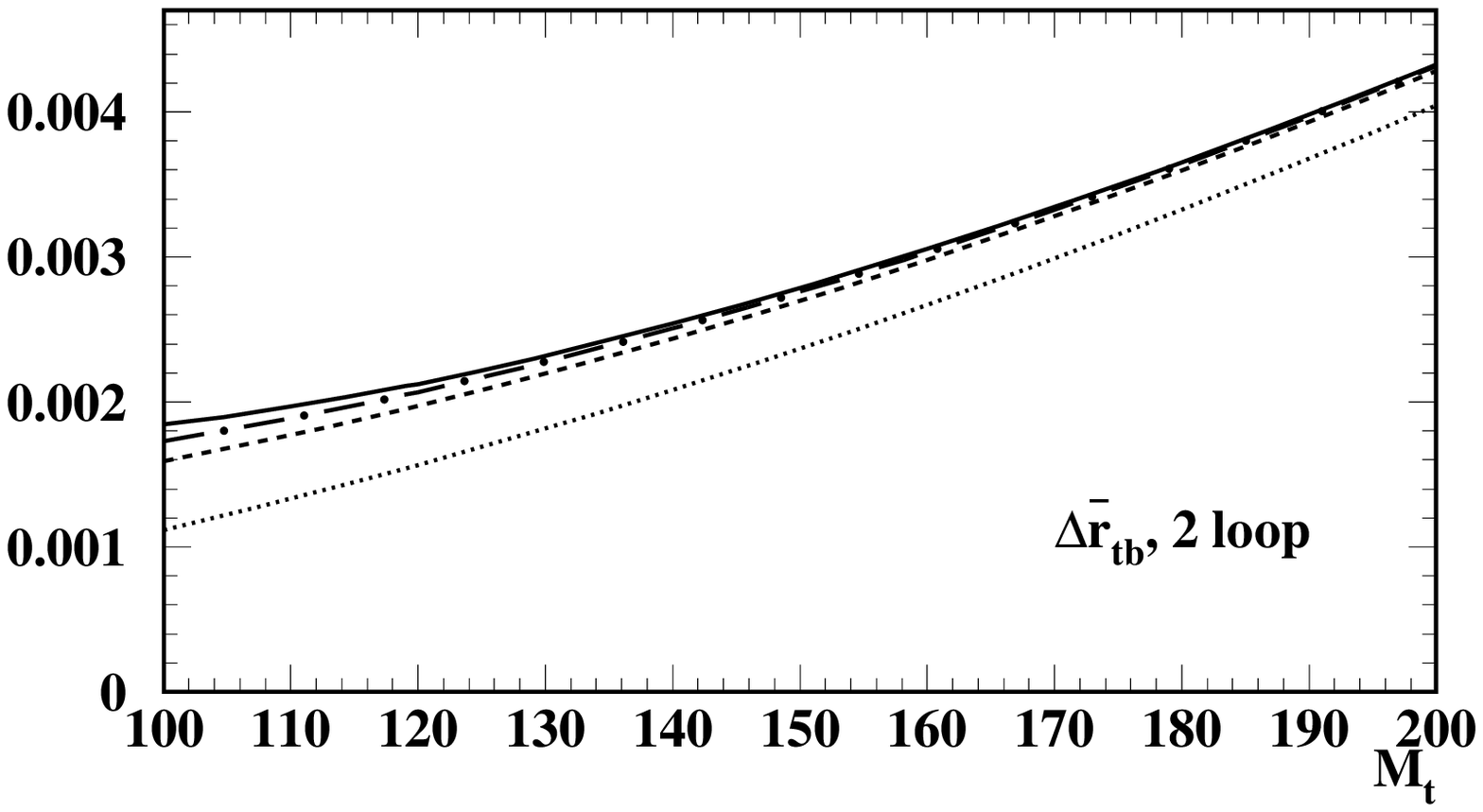} \\
   \epsfxsize=11.5cm
   \leavevmode
   \epsffile[50 290 520 560]{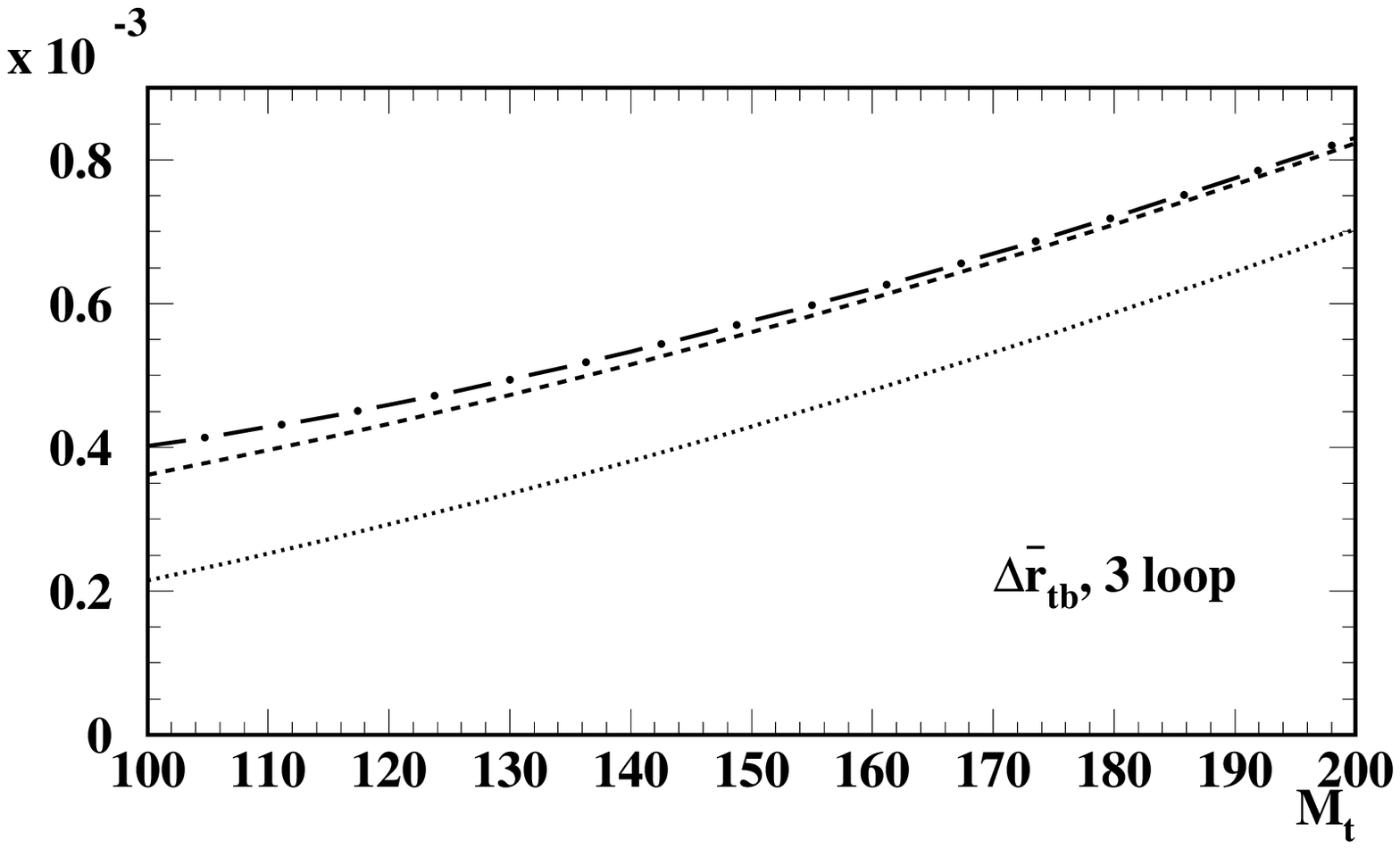}
 \end{tabular}
 \caption{}
 \end{center}
\end{figure}

\begin{figure}[ht]
 \begin{center}
 \begin{tabular}{c}
   \epsfxsize=11.5cm
   \leavevmode
   \epsffile[50 290 520 560]{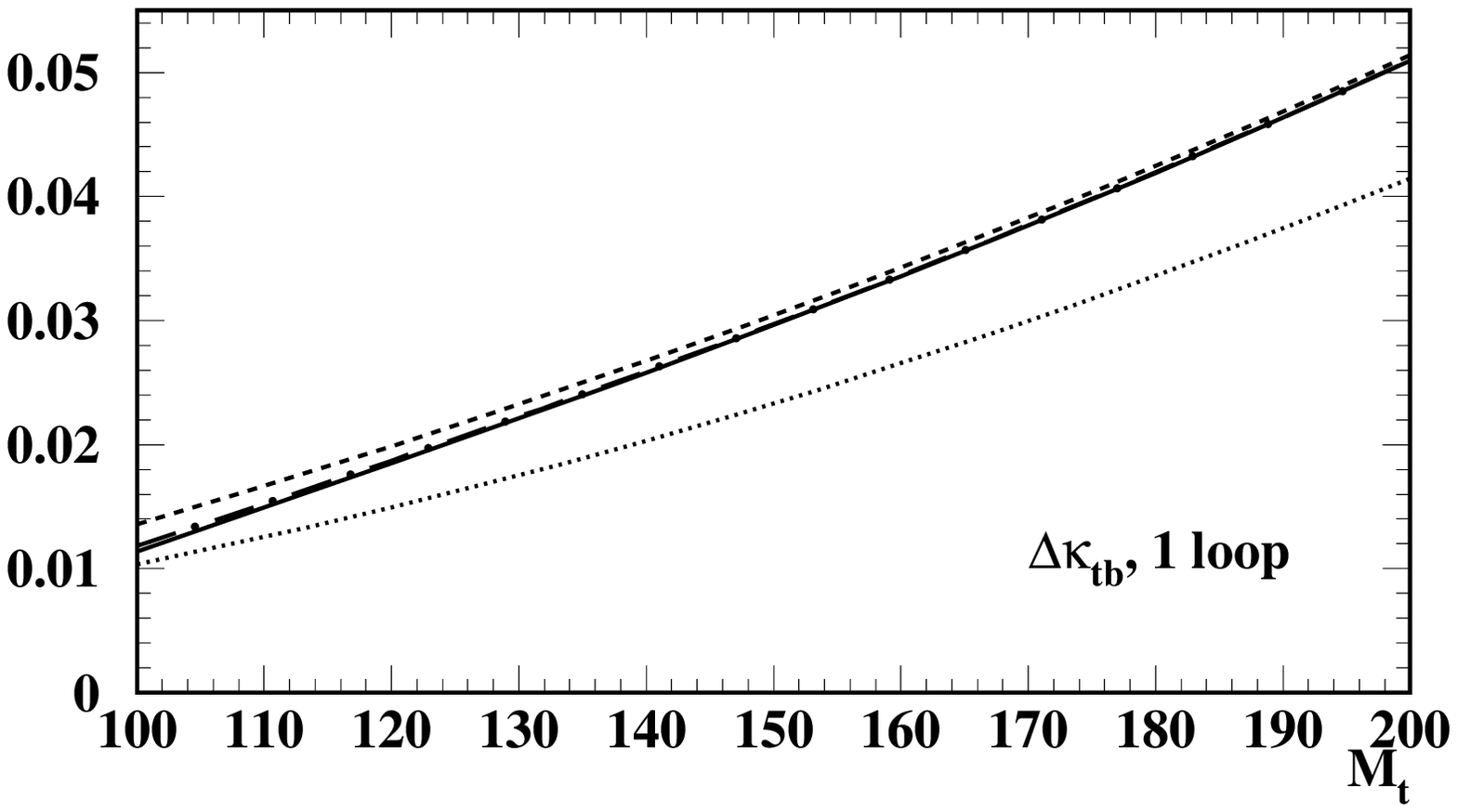} \\
   \epsfxsize=11.5cm
   \leavevmode
   \epsffile[50 290 520 560]{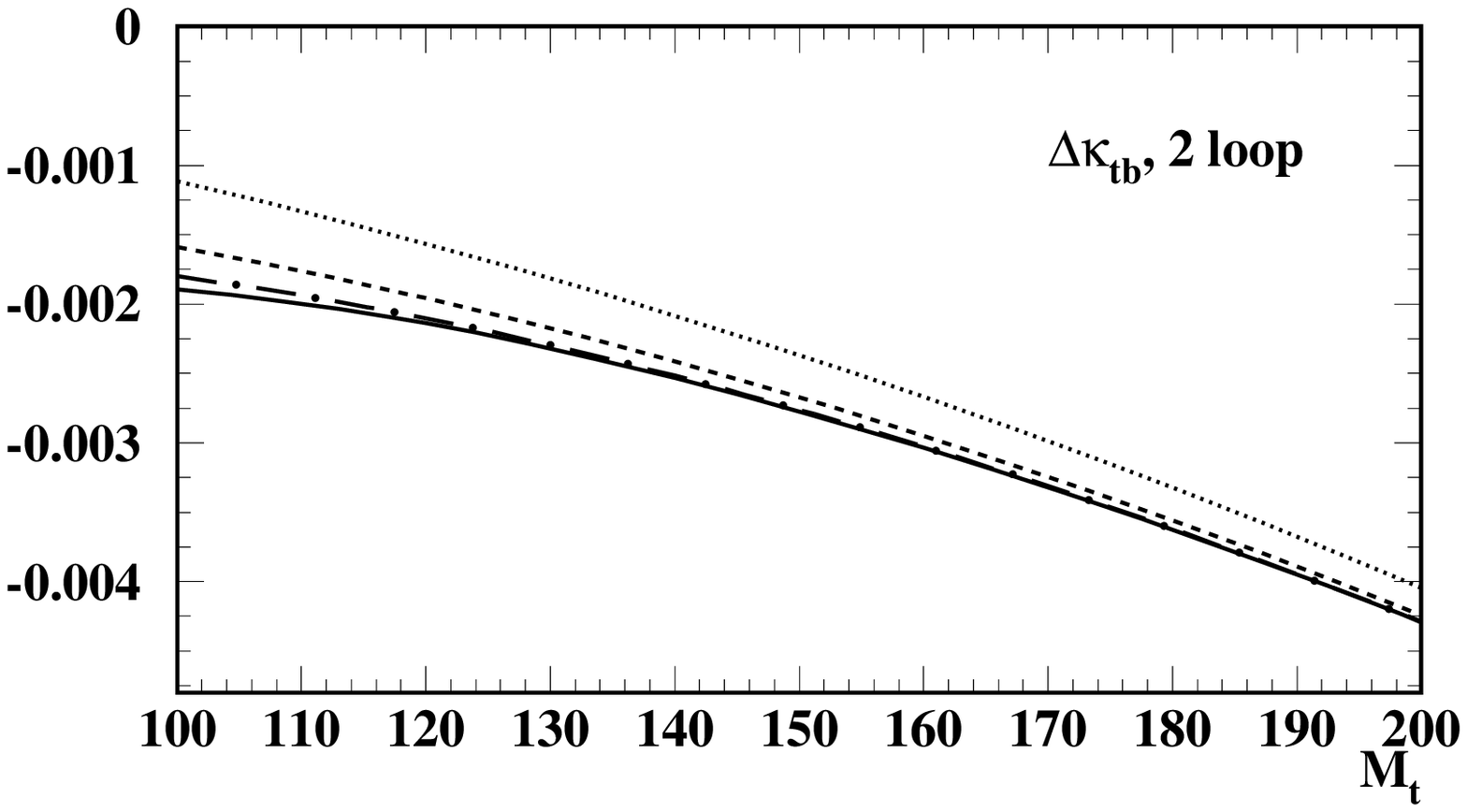} \\
   \epsfxsize=11.5cm
   \leavevmode
   \epsffile[50 290 520 560]{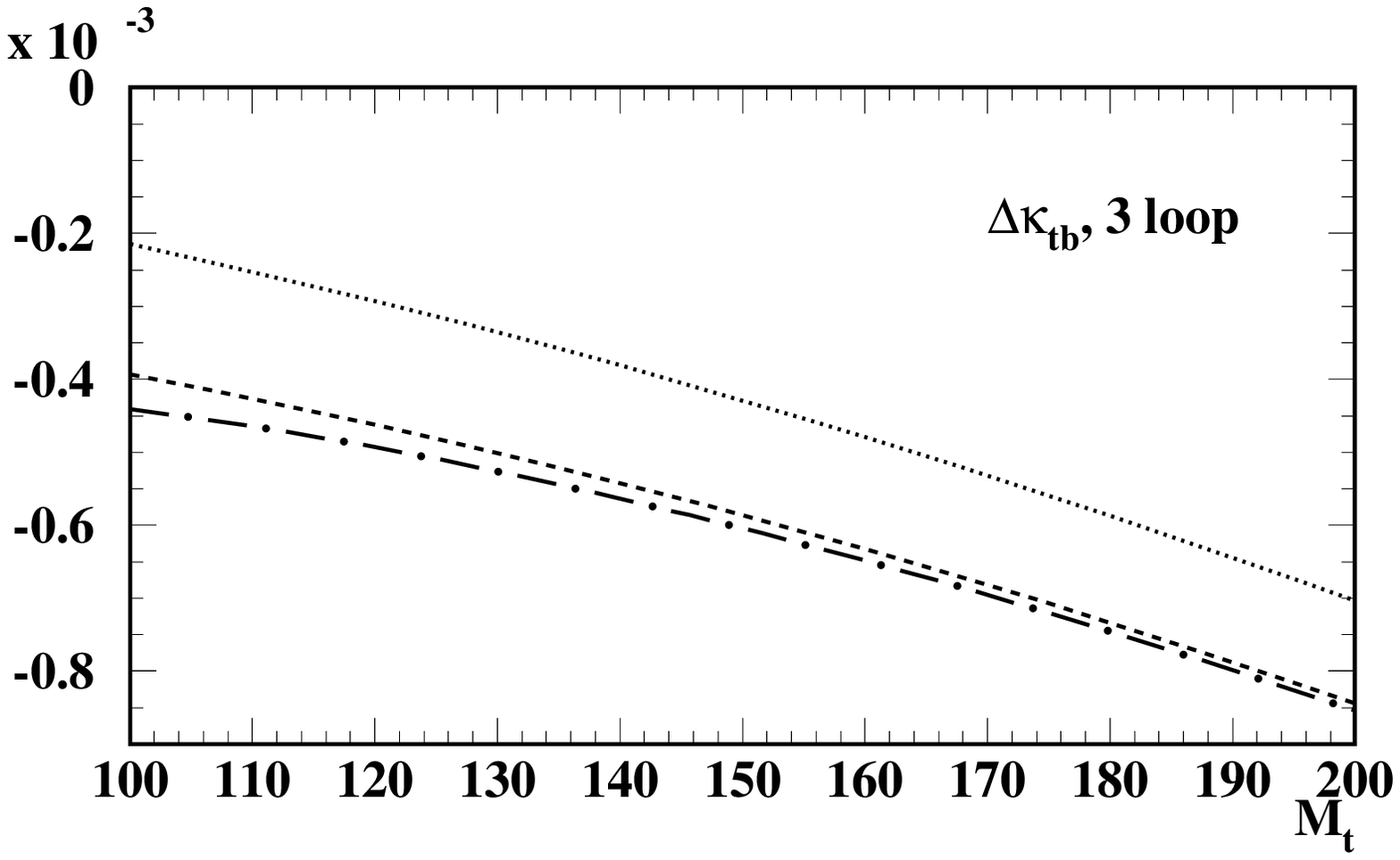}
 \end{tabular}
  \caption{}
 \end{center}
\end{figure}

\end{document}